\documentclass[prx,twocolumn,aps,showpacs,amsmath,amssymb]{revtex4-1}

\usepackage{graphicx}
\usepackage{epstopdf}
\usepackage{dcolumn}
\usepackage{bm}
\usepackage{amsmath}

\begin{document}




\title{Light Quantum Control of Persisting Higgs Modes in Iron-Based Superconductors} 


\author
{C. Vaswani$^{1\ast}$, J.~H.~Kang$^{2\ast}$, M.~Mootz$^{3\ast}$, L. Luo$^{1}$, X. Yang$^{1}$, C.~Sundahl$^{2}$, D. Cheng$^{1}$, C. Huang$^{1}$, R.~H.~J.~Kim$^{1}$, Z. Liu$^{1}$, Y. G. Collantes$^{4}$, E. E. Hellstrom$^{4}$, I.~E.~Perakis$^{3}$, C.~B.~Eom$^{2}$ and J. Wang$^{1\dag}$}

\affiliation{$^1$Department of Physics and Astronomy, Iowa State University, and Ames Laboratory, Ames, IA 50011 USA.
	\\$^2$Department of Materials Science and Engineering, University of Wisconsin-Madison, Madison, WI 53706, USA.
	\\$^3$Department of Physics, University of Alabama at Birmingham, Birmingham, AL 35294-1170, USA.
\\$^4$Applied Superconductivity Center, National High Magnetic Field Laboratory, Florida State University, Tallahassee, FL 32310, USA.}
\date{\today}

\begin{abstract}
	The Higgs mechanism, i.e., spontaneous symmetry breaking of the quantum vacuum, is a cross-disciplinary principle, universal for understanding dark energy, antimatter and quantum materials, from superconductivity to magnetism.  
	Yet,  Higgs modes in one-band superconductors (SCs) are currently under debate
	due to their competition with charge-density fluctuations.
	A  distinct Higgs mode, controllable by terahertz (THz) laser pulses, can  arise in  multi-band, unconventional SCs via strong {\em interband} Coulomb interaction, but is yet to be accessed. 
	Here we both discover and demonstrate quantum control of such 
	collective  mode in iron-based high-temperature superconductors. 
	Using two-pulse, phase coherent THz spectroscopy,
	we observe a tunable and coherent 2$\Delta_{\mathrm{SC}}$  amplitude oscillation of the complex order parameter in such SC with coupled lower and upper bands.
	The nonlinear dependence  of the amplitude mode oscillations on the THz driving fields is distinct from any one-band and conventional SC results: we observe a large nonlinear change  of resonance strength, yet with a persisting mode frequency. 
	We argue that this  result provides compelling evidence for 
	a transient coupling 
	between 
	the electron and hole amplitude modes via strong interband 
	coherent interaction. To  support  this scenario, we perform quantum kinetic modeling of a {\em hybrid} Higgs mechanism without invoking extra disorder or phonons.
	In addition to distinguishing between collective modes and charge fluctuations, 
	the light quantum control of multiband SCs
	can be extended to probe and manipulate many-body entanglement and hidden symmetries in different quantum materials.
\end{abstract}

\maketitle

\maketitle

Phase coherence between multiple SC condensates in different strongly interacting bands is well--established in iron-based superconductors (FeSCs). As illustrated in Fig.~1a, a dominant Coulomb coupling between the $h$- and $e$-like Fermi sea pockets, unlike in other SCs, is manifested by, e.g., $s\pm$ pairing symmetry \cite{Chubukov:2012,Chubukov:2015}, spin density wave resonant peaks and nesting wave vectors (black arrows) \cite{ref5, 1,2,ref4}.
Experimental evidence for Higgs amplitude coherent excitations in FeSCs has not been reported yet, despite recent progress in non-equilibrium superconductivity and collective modes~\cite{giorgianni2019leggett, Matsunaga:2013, matsunaga2014,Cea2016,Rajasekaran, n1, n2, n3, n4,Wu2019,Udina:2019,Kumar:2019,Manske:2020,Kaiser2020,Podolsky2020}.

The condensates in different bands of Ba(Fe$_{1-x}$Co$_{x}$)$_2$As$_2$ studied here, shown in Fig.~1a, are coupled by the strong interband $e$--$h$ interaction $U$ (blue double arrow vector), which is about one order of magnitude stronger than the intraband interaction $V$ (gray and red thin lines)~\cite{Fernandes:2010,Akbari:2013}. 
For $U\gg V$, the formation channels of collective modes are distinct from one-band SC~\cite{matsunaga2014,Cea2016,Cea2018,Udina:2019} and multiband MgB$_2$ with dominant {\em intraband} interaction, $V\gg U$~\cite{Shimano2019, Aoki2017,krull2016}. For the latter, only Leggett modes are observed thus far \cite{giorgianni2019leggett, ref1}. 
In contrast, for $U\gg V$, one expects Higgs amplitude modes arising from the condensates in all Coulomb--coupled bands, i.e., in the $h$ pocket at the $\Gamma$-point (gray circle, mode frequency $\omega_{\mathrm{H,1}}$), and in the two $e$ pockets at (0, $\pi$) and ($\pi$, 0) (red ellipses, $\omega_{\mathrm{H,2}}$). 
A single-cycle THz oscillating field 
(red pulse) can act like a ``quantum quench", with impulsive non-adiabatic driving of the Mexican-hat-like quantum fields (dark green) and, yet, with minimum heating of other degrees of freedoms.
Consequently, the multi-band condensates are forced out of the free energy minima, since they cannot follow the quench adiabatically. Most intriguingly, such coherent nonlinear driving not only excites amplitude mode oscillations in the different Fermi sea pockets, but also transiently modifies their coupling, assisted by the strong interband interaction $U$. Such coherent transient coupling can be regarded as nonlinear amplitude mode hybridization with a time-dependent phase coherence. In this way, THz laser fields can manipulate phase-coherent, {\em hybrid} Higgs emerging collective modes.


\begin{figure*}[!tbp]
	\includegraphics[scale=0.6]{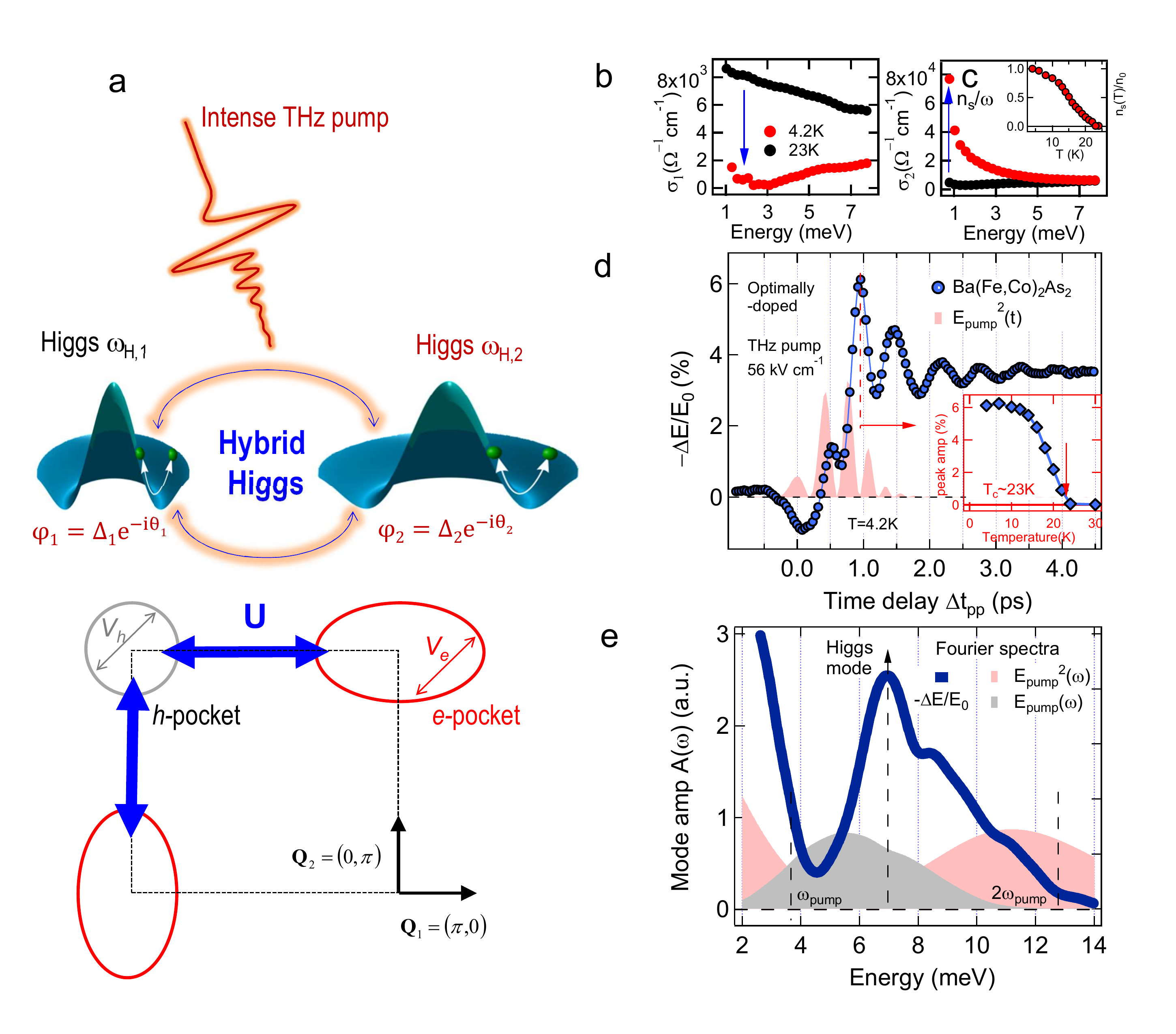}
	\caption{\textbf{The 2$\Delta_{\mathrm{SC}}$ oscillations detected by two-pulse THz coherent spectroscopy of multi-band FeSCs.} 
		$\textbf {a},$ Illustration of coherent excitation of hybrid Higgs mode via THz quantum quench. An effective three-band model has a $h$ pocket at the $\Gamma$ point and two $e$ pockets at X/Y points, with strong inter- (blue) and weak intra-band (gray) interactions marked by arrows. 
		$\textbf {b,c,}$ Real and imaginary parts of the complex THz conductivity spectra $\sigma_1(\omega)$ and $\sigma_2(\omega)$ in the superconducting (4.2~K) and normal states (23~K) of Ba(Fe$_{1-x}$Co$_{x}$)$_2$As$_2$ (x=0.08) in equilibrium. Inset of \textbf{c} shows the temperature dependence of the superfluid density normalized to its value at 4.2K, n$_s(T)$/n$_0$, as determined from $1/\omega$ divergence of $\sigma_2(\omega)$ (blue arrow in $\textbf c$).
		$\textbf d,$ Differential THz transmission $\Delta E/E_{0}$ (blue circles) measured by phase-locked two-THz-pulse pump--probe spectroscopy at 4.2~K shows pronounced coherent oscillations for a peak THz field strength of E$_{\mathrm{pump}}$=56~kV/cm. The pink shaded curve denotes the square of the pump THz waveform E$_{\mathrm{pump}}^2$. Inset: Temperature dependence of the peak amplitude of $\Delta E/E_{0}$. 
		$\textbf {e},$ Fourier spectrum of the coherent oscillations in $\Delta E_{\mathrm{}}/E_{0}$ (blue line) exhibits a resonance peak at $\sim$6.9~meV (blue line) and is distinct from both the pump E$_{\mathrm{pump}}$ (gray) and pump-squared E$_{\mathrm{pump}}^2$ (pink) spectra.}
	\label{Fig1}
\end{figure*}

Here we present evidence of 
Higgs modes that are controlled by THz-field-driven interband quantum entanglement in a multi-band SC, optimally-doped Ba(Fe$_{1-x}$Co$_{x}$)$_2$As$_2$, 
using two phase-locked near-single–cycle THz laser fields.   
We thus reveal a striking nonlinear THz field dependence of coherent amplitude mode oscillations: quick increase to maximum spectral weight (SW) with negligible mode frequency shift, followed by a huge SW reduction by more than 50$\%$, yet with robust mode frequency position, with less than 10$\%$ redshift.   
These distinguishing features of the 
 observed collective mode are different from any one-band and conventional SC results and predictions so far. Instead, they are predicted by our quantum kinetic calculation, which identifies the key role of the interband interaction $U$ for coherently coupling two amplitude modes, in the $h$- and $e$-like Fermi sea pockets, and for controlling the SW of the lower Higgs mode observed in the experiment.  

The equilibrium complex conductivity spectra, i.e., real and imaginary parts,  ${\sigma}_1 (\omega)$ and ${\sigma}_2 (\omega)$, of our epitaxial Ba(Fe$_{1-x}$Co$_{x}$)$_2$As$_2$ (x=0.08) film~\cite{lee2010} (Methods) measure the low-frequency quasi-particle (QP) electrodynamics and condensate coherence, respectively (Figs. 1b and 1c)~\cite{Xu}. 
The normal state (black circles) displays Drude--like behavior, while the QP spectral weight in ${\sigma}_1(\omega)$ is depleted in the SC state due to SC gap openings, seen, e.g., in the 4.2~K trace (red circles). 
The lowest SC gap value 2$\Delta_{1}\sim$6.8 meV obtained is in agreement with the literature values 6.2-7 meV \cite{ref2,ref3} (Methods).  
Such ${\sigma}_1 (\omega)$ spectral weight depletion is accompanied by an increase of condensate fraction $n_\mathrm{s}/n_\mathrm{0}$ (inset, Fig.~1c), extracted from a diverging $1/\omega$ condensate inductive response, marked by blue arrow, e.~g., in the 4.2~K lineshape of ${\sigma}_2(\omega)$ (Fig.~1c). Note that superfluid density $n_\mathrm{s}$ vanishes above T$_c \sim$23\,~K (inset Fig.~1c).      
     
We characterize the THz quantum quench coherent dynamics directly in the time domain (Methods) by measuring the pump--probe responses to two phase-locked THz pulses as differential field transmission $\Delta E_{\mathrm{}}/E_{0}$ (blue circles, Fig.~1d) for 
THz pump field, $E_{\mathrm{pump}}=$56~kV/cm as a function of pump--probe time delay $\Delta t_\mathrm{pp}$. 
The  central pump energy $\hbar \omega_\mathrm{pump}=$5.4~meV (gray shade, Fig.~1e) is chosen slightly below the 
2$\Delta_{1}$ gap.
Intriguingly, the  $\Delta E/E_{0}$ dynamics reveals a pronounced coherent oscillation, superimposed on the overall amplitude change, which persists much longer than the THz photoexcitation (pink shade). 
This mode is excited by the quadratic coupling of the pump vector potential, $\mathbf{A}^{2}(t)\propto E^2_{\mathrm{pump}}/\omega^2$ due to the SC equilibrium symmetry~\cite{matsunaga2014}.  
Such coherent responses yield information within the general framework of 2D coherent nonlinear spectroscopy~\cite{rupert} (Methods). 
The origin of the observed coherent $\Delta E/E_{0}$ oscillation is better illustrated by its Fourier transformation (FT), shown in Fig.~1e. The FT spectrum of the coherent nonlinear signals (blue solid line) displays a {\em pronounced} resonance at 6.9meV, indicative of 2$\Delta_{1}$ coherent amplitude mode oscillations. This FT spectrum strongly differs from the spectra of both THz pump E$_{\mathrm{pump}}(\omega)$ centered at $\omega_{{\mathrm{pump}}}\sim$5.4 meV (gray shade) and second harmonic, Anderson pseudo-spin (APS) precession at 2$\omega_{\mathrm{pump}}$ from E$^2_{\mathrm{pump}}(\omega)$ (pink shade). This is a consequence of the broadband spectrum of the few-cycle pump pulse used in the experiment which overlaps with the mode resonances such that $\Delta E/E_{0}$ oscillates with the collective mode frequencies~\cite{Udina:2019}.   
After the oscillation, time-dependent complex conductivity spectra, $\sigma_1(\omega, \Delta t_{\mathrm{pp}})$ and $\sigma_2(\omega, \Delta t_{\mathrm{pp}})$, can be measured (Figs.S5-S6, Supplementary). They show that $\Delta E/E_{0}$ closely follows the pump-induced change in condensate density, $\Delta n_\mathrm{s}/n_\mathrm{0}$~\cite{n1}. The THz excitation at $E_{\mathrm{pump}}=$56 kV/cm 
only reduces $n_\mathrm{s}$ slightly, $\Delta n_\mathrm{s}/n_\mathrm{0}\sim \Delta E_{\mathrm{}}/E_{0} \sim -3~\%$ at $\Delta t_\mathrm{pp}=5$~ps. 
Furthermore, the pump-induced peak amplitude (blue diamond), marked by the red dashed line in Fig.~1d, diminishes above T$_c$ (inset).
The measured coherent oscillations reflect the emergence of a hybrid Higgs multi-band collective mode between two Coulomb–coupled lower and higher modes, $\omega_{\mathrm{H,1}}$ and $\omega_{\mathrm{H,2}}$, as shown later. 

\textbf{Fig. 1. The 2$\Delta_{\mathrm{SC}}$ oscillations detected by two-pulse THz coherent spectroscopy of multi-band FeSCs.} 
$\textbf {a},$ Illustration of coherent excitation of hybrid Higgs mode via THz quantum quench. An effective three-band model has a $h$ pocket at the $\Gamma$ point and two $e$ pockets at X/Y points, with strong inter- (blue) and weak intra-band (gray) interactions marked by arrows. 
$\textbf {b,c,}$ Real and imaginary parts of the complex THz conductivity spectra $\sigma_1(\omega)$ and $\sigma_2(\omega)$ in the superconducting (4.2~K) and normal states (23~K) of Ba(Fe$_{1-x}$Co$_{x}$)$_2$As$_2$ (x=0.08) in equilibrium. Inset of \textbf{c} shows the temperature dependence of the superfluid density normalized to its value at 4.2K, n$_s(T)$/n$_0$, as determined from $1/\omega$ divergence of $\sigma_2(\omega)$ (blue arrow in $\textbf c$).
$\textbf d,$ Differential THz transmission $\Delta E/E_{0}$ (blue circles) measured by phase-locked two-THz-pulse pump--probe spectroscopy at 4.2~K shows pronounced coherent oscillations for a peak THz field strength of E$_{\mathrm{pump}}$=56~kV/cm. The pink shaded curve denotes the square of the pump THz waveform E$_{\mathrm{pump}}^2$. Inset: Temperature dependence of the peak amplitude of $\Delta E/E_{0}$. 
$\textbf {e},$ Fourier spectrum of the coherent oscillations in $\Delta E_{\mathrm{}}/E_{0}$ (blue line) exhibits a resonance peak at $\sim$6.9~meV (blue line) and is distinct from both the pump E$_{\mathrm{pump}}$ (gray) and pump-squared E$_{\mathrm{pump}}^2$ (pink) spectra.

\begin{figure*}[!tbp]
	\includegraphics[scale=0.6]{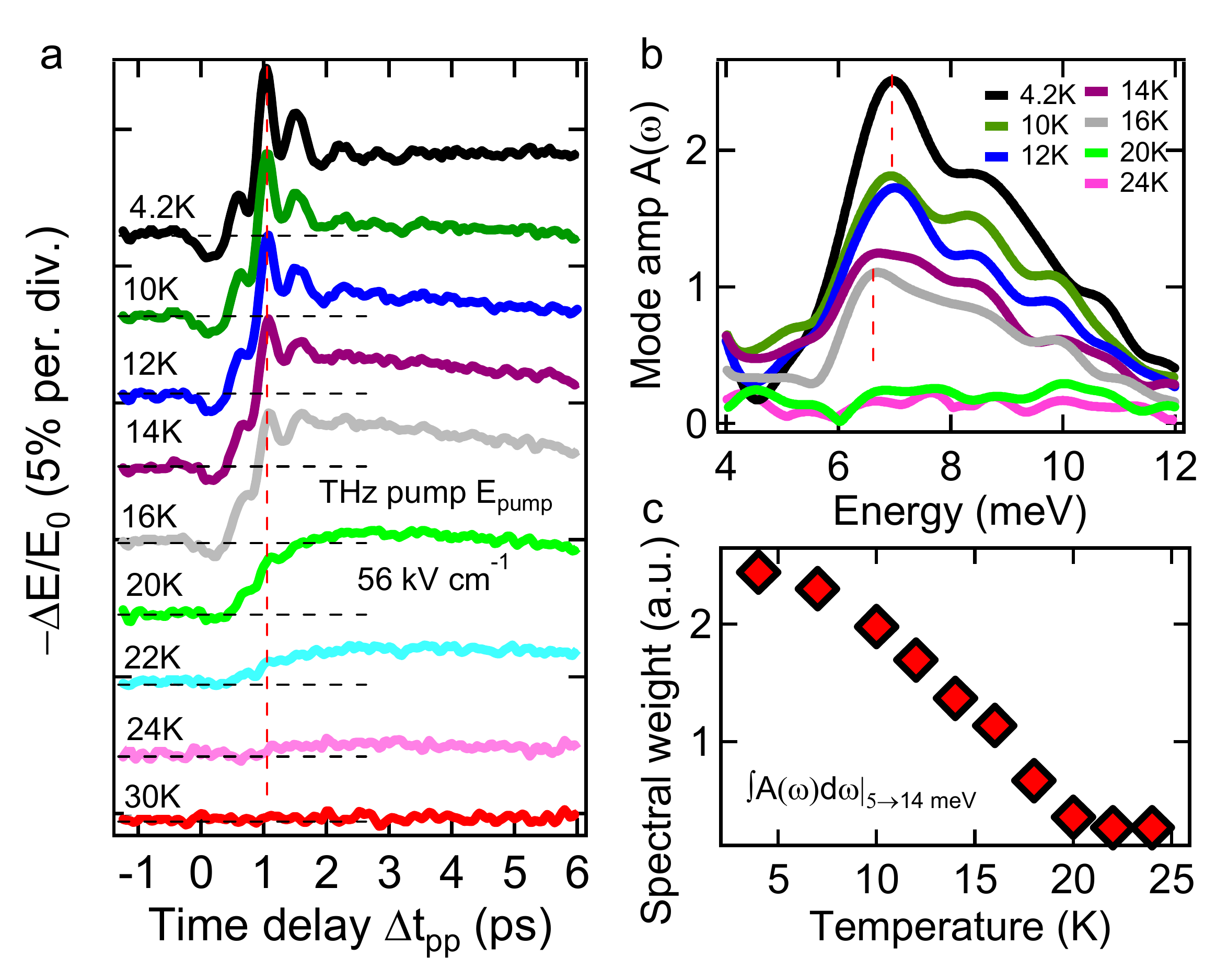}
	\caption{\textbf{Temperature dependence of hybrid Higgs coherent oscillations in FeSCs.} 
		$\textbf {a},$ Temporal profiles of $-\Delta E_{\mathrm{}}/E_{0}$ at various temperatures and for a peak THz pump E-field of E$_{\mathrm{pump}}$=56~kV/cm. Traces are offset for clarity. 
		$\textbf {b},$ Fourier spectra of the Higgs mode oscillations derived from the two-pulse coherent pump--probe signals at different temperatures. 
		Dashed red lines indicate the slight redshift of the Higgs mode frequency with the drastic reduction of mode SW as a function of temperature. 
		$\textbf {c},$ Temperature dependence of the integrated SW$_{5\rightarrow14~\mathrm{meV}}$ of the Higgs mode in \textbf{b}.}
	\label{Fig1}
\end{figure*}

Figure~2 reveals a {\em strong temperature dependence} of the Higgs oscillations. 
The coherent dynamics of $\Delta E/E_{0}$ is shown in Fig.~2a for temperatures 4.2~K--30~K.  
Approaching $T_c$ from below, the coherent oscillations quickly diminish, as seen by comparing the 4.2~K (black line) and 16~K (gray) traces versus 22~K (cyan) and 24~K (pink) traces.  
Fig.~2b shows the temperature--dependent Fourier spectra of $\Delta E/E_{0}$, in the range 4--12~meV. 
Fig.~2c plots the integrated spectral weight SW$_{5\rightarrow14~\mathrm{meV}}$ of the amplitude mode (Fig.~2b).    
The strong temperature dependence correlates the mode with SC coherence. Importantly, while the mode frequency is only slightly red-shifted, by less than 10~$\%$ before full SW depletion close to $T_c$, SW is strongly suppressed, by $\sim$55~$\%$ at 16~K ($T/T_c\sim$0.7).
Such a spectral weight reduction in FeSCs is much larger than that expected in one-band BCS superconductors or for $U$=0 shown later (Figs. 4e-4f). 
We also note that the temperature dependence of Higgs oscillations, observed in FeSCs here, has not been measured experimentally in conventional BCS systems, and could represent a key signature of quantum quench dynamics of unconventional SCs. 
Note that the observed behavior is 
consistent with our simulations of the
Higgs mode in multi-band SCs with dominant interband $U$
, shown later in Figs.~4e-4f.

Figure~3 presents the distinguishing experimental evidence for the unconventional 
Higgs mode in FeSCs, which is different from one-band SCs -- a highly nonlinear THz electric field dependence of coherent $2\Delta_{1}$ oscillations that manifests as a huge SW change, yet with persisting mode frequency, with only very small redshift.  
Fig.~3a shows the detailed pump-fluence dependence of $\Delta E_{\mathrm{}}/E_{0}$ as a function of time delay, presented as a false-color plot at $T$=4.2\,~K for up to E$_{\mathrm{pump}} \sim 600$~kV/cm.  
It is clearly seen that amplitude mode oscillations depend nonlinearly on the $E_{\mathrm{pump}}$ field strength. 
1$+\Delta E_{\mathrm{}}/E_{0}$ (red solid line) at $\Delta t_\mathrm{pp}=5$~ps is shown in Fig.~3b. This, together with the measured $\sim 1/\omega$ divergence in $\sigma_{2}(\omega, \Delta t_\mathrm{pp})$, allows the determination of the condensate fraction $n_\mathrm{s}(E_\mathrm{pump})/n_\mathrm{0}$ (blue circles) in the driven state (Fig. S6, Supplementary).
This shows three different excitation regimes, marked by black dash lines in Figs.~3a-3b: (1) in regime \#1, the condensate quench is {\em minimal}, e.~g., $n_\mathrm{s}/n_\mathrm{0}\geq$97\% below the field $E_{\#1}=56$~kV/cm; (2) Regime \#2 displays {\em partial SC quench}, where $n_\mathrm{s}/n_\mathrm{0}$ is still significant, e.~g., condensate fraction $\approx$60\% at $E_{\#2}=276$~kV/cm; (3) A {\em saturation} regime \#3 is observed $\sim E_{\#3}=600$~kV/cm, which leads to a slowly changing $n_\mathrm{s}/n_\mathrm{0}$ approaching 25~\%. The saturation is expected since the below gap THz pump is used, especially $\hbar\omega_{pump}\ll$2$\Delta_{2}\sim$15-19 meV at the $e$ pockets~\cite{ref2, ref3}. 

\begin{figure*}[!tbp]

	\includegraphics[scale=0.5]{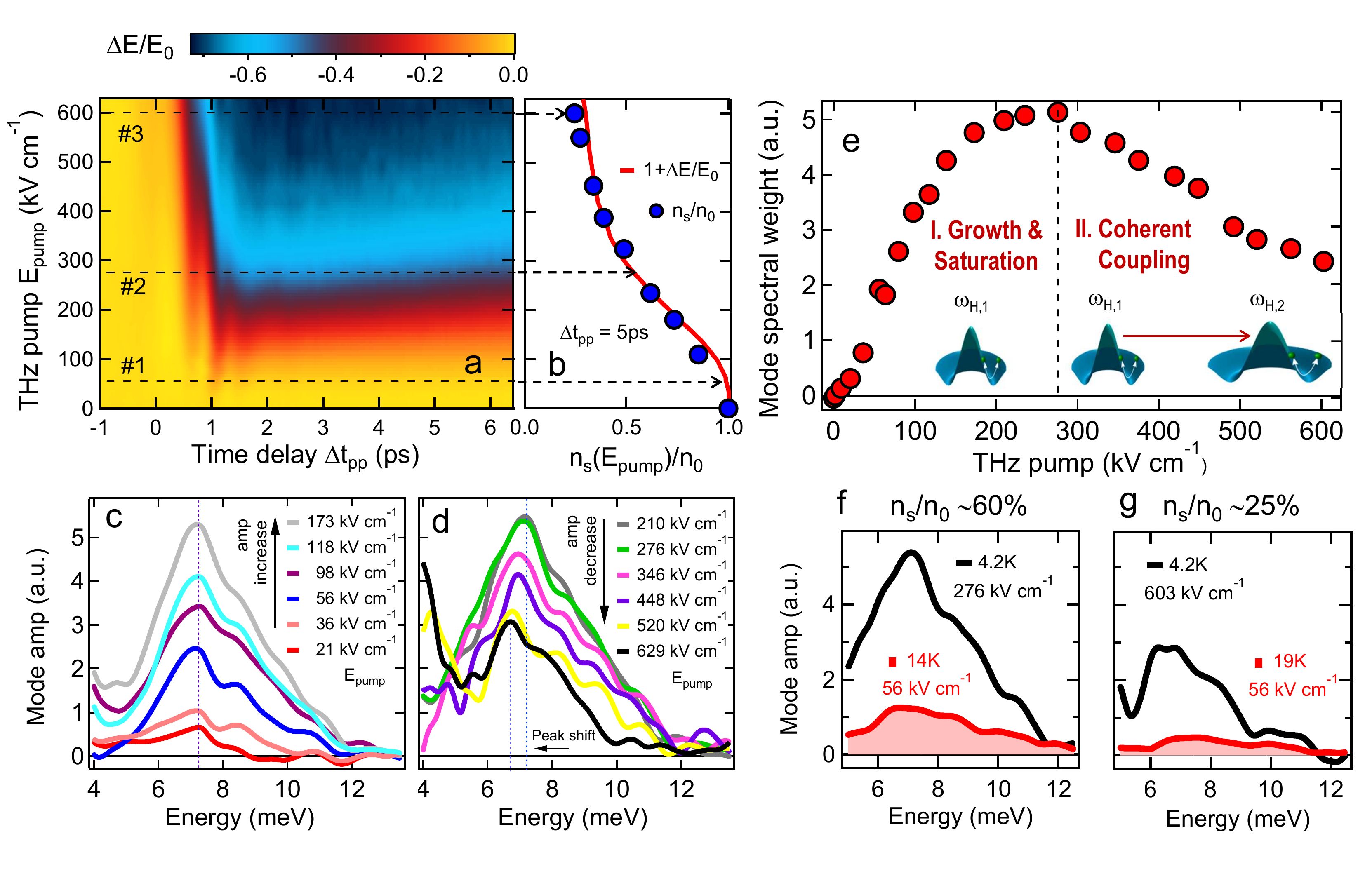}
	\caption{\textbf{Nonlinear THz field dependence of spectral weight of
			Higgs mode in FeSCs.} $\textbf {a},$ A 2D false-color plot of $\Delta E_{\mathrm{}}/E_{0}$ as a function of pump E-field strength E$_{\mathrm{pump}}$ and pump--probe delay $\Delta t_\mathrm{pp}$ at 4.2~K. 	$\textbf {b},$ THz pump $E$-field E$_{\mathrm{pump}}$ dependence of $1+\Delta E_{\mathrm{}}/E_{0}$ (red line) overlaid with the superfluid density fraction n$_s$/n$_0$ (blue circles) after THz pump at $\Delta t_\mathrm{pp}$=5ps and T=4.2~K. Dashed arrows mark the three pump $E$-field regimes, i.~e., {\em weak}, {\em partial}, and {\em saturation}, identified in the main text. 
	$\textbf {c, d}$ Spectra of coherent Higgs mode oscillations show a distinct non-monotonic dependence as a function of THz pump field, i.e., a rapid increase in the mode amplitude for low pump E-field strengths up to 173kV/cm, saturation up to 276kV/cm and significant reduction at higher fields. The blue dashed line marks the resonance of the mode and the redshift of the Higgs mode peak, much smaller than the mode SW change.
	$\textbf {e},$ Integrated spectral weight SW$_{5\rightarrow14~\mathrm{meV}}$ of the Higgs mode at various pump $E$-field strengths, indicating the SW reduction of the Higgs mode from dominant $\omega_{\mathrm{\mathrm{H},1}}$ at low driving fields to $\omega_{\mathrm{\mathrm{H},2}}$ due to the interband interaction and coherent coupling at high driving fields above $E_{\mathrm{pump}}=$276~kV/cm (inset). 
	$\textbf {f, g}$ Contrasting the thermal and THz-driven states of coherent hybrid Higgs responses by comparing the mode spectra at the similar superfluid density \textbf{f} $n_s/n_0 = 60\%$ and \textbf{g} $n_s/n_0 = 25\%$ achieved by THz pump (black solid lines) and temperature (red shades).}
	\label{Fig1}
\end{figure*}

Quantum quenching of the single-band BCS pairing interaction has been well established to induce Higgs oscillations with amplitude scaling as 1/$\sqrt{\Delta_{SC, \infty}}$, determined by the long--time asymptotic nonthermal order parameter $\Delta_{SC, \infty}$. The latter 
 decreases with pump field~\cite{Yuzbashyan:2006,Axt:2007,Forster:2017}. Both model and experimental results establish that the Higgs mode amplitude increases with THz pumping until full depletion of the condensate, concurrent with a  continuous Higgs resonance redshift to zero~\cite{Yuzbashyan:2006,Axt:2007,Forster:2017, Matsunaga:2013}.
In contrast to this expected behavior for conventional SCs, Fig.~3a and the Fourier spectra of the coherent oscillations, Figs.~3c and 3d, show a non-monotonic pump-field dependence of the Higgs mode amplitude
that is unique here. 
Specifically, the Fourier spectra exhibit a clear resonance at low pump fluences, which coincides with the frequency of the lower Higgs mode $\omega_{\mathrm{\mathrm{H},1}}$. 
This resonance grows quickly up to a field of $E_\mathrm{pump}=173$~kV/cm (Fig.~3c), saturates up to 276~kV/cm (Fig.~3d) and then exhibits a significant reduction in pump regime \#3, e.g., by more than 50~$\%$ at 629~kV/cm (black line, Fig.~3d). 
This non-monotonic field dependence clearly shows that the coherent oscillations in Fig. 3a quickly increase in pump regime \#1 and saturate in regime \#2, prior to any significant mode resonance redshift (blue dashed line, Fig. 3c). 
Above this relatively low field regime, the oscillation amplitude starts to decrease at 276~kV/cm, even though there is still more than 60\% of  condensate, marked in Fig.~3b:  the driven state is still far from full SC depletion.
This striking SW reduction is also seen in the integrated spectral weight analysis, SW$_{5\rightarrow14~\mathrm{meV}}$, summarized in Fig.~3e. 
Most intriguingly, while there is a large reduction of the Higgs mode SW$\sim$50\% at 629~kV/cm (regime \#3), the resonance peak position is nearly persistent, with $\leq$10~$\%$ red shift (blue dashed lines, Fig. 3d). 
These observations differ from any known behavior in one-band SCs, but are consistent with expectations from coherent coupling of Higgs modes in multi-band SCs by a dominant interband interaction $U$, discussed below.  

The distinct mode amplitude and position variation with pump field extracted from the coherent oscillations 
clearly show the transition from SW growth and saturation to reduction, marked by the black dashed line at $E_\mathrm{pump}=$276~kV/cm (Fig. 3e). The saturation and reduction of SW in the amplitude oscillation, yet with a persisting mode frequency, cannot be explained by any known mechanism. This can arise from the coupling of the two amplitude modes $\omega_{\mathrm{\mathrm{H},1}}$ and $\omega_{\mathrm{\mathrm{H},2}}$ 
expected in iron pnictides 
 due to the strong inter-pocket interaction $U$. As we demonstrate theoretically below,  the coherent coupling process shown in Fig. 3e  can be controlled and detected nonlinearly by two phase-locked THz pulses.
 We argue that  (I) At low driving fields, $\omega_{\mathrm{\mathrm{H},1}}$ dominates the hybrid collective mode due to less damping than $\omega_{\mathrm{\mathrm{H},2}}$ arising from the asymmetry between the electron and hole pockets; (II) For higher fields, SW of $\omega_{\mathrm{\mathrm{H},1}}$ mode decreases due to the strong coupling with $\omega_{\mathrm{\mathrm{H},2}}$ expected for the strong inter--band interaction in iron pnictides. Moreover, it is critical to note that the THz driving is of highly non-thermal nature, which is distinctly different from that obtained by temperature tuning in Fig.~2. Specifically, Figs.~3f and 3g compare the hybrid Higgs mode spectra for similar condensate faction $n_\mathrm{s}/n_\mathrm{0}$, i.e., $\approx$60\% (f) and 25\% (g), induced by tuning either the temperature (red shade) or THz pump (black line). The mode is clearly much stronger in the THz driven coherent states than in the temperature tuned ones.      

\begin{figure*}[!tbp]
	
	\includegraphics[scale=0.8]{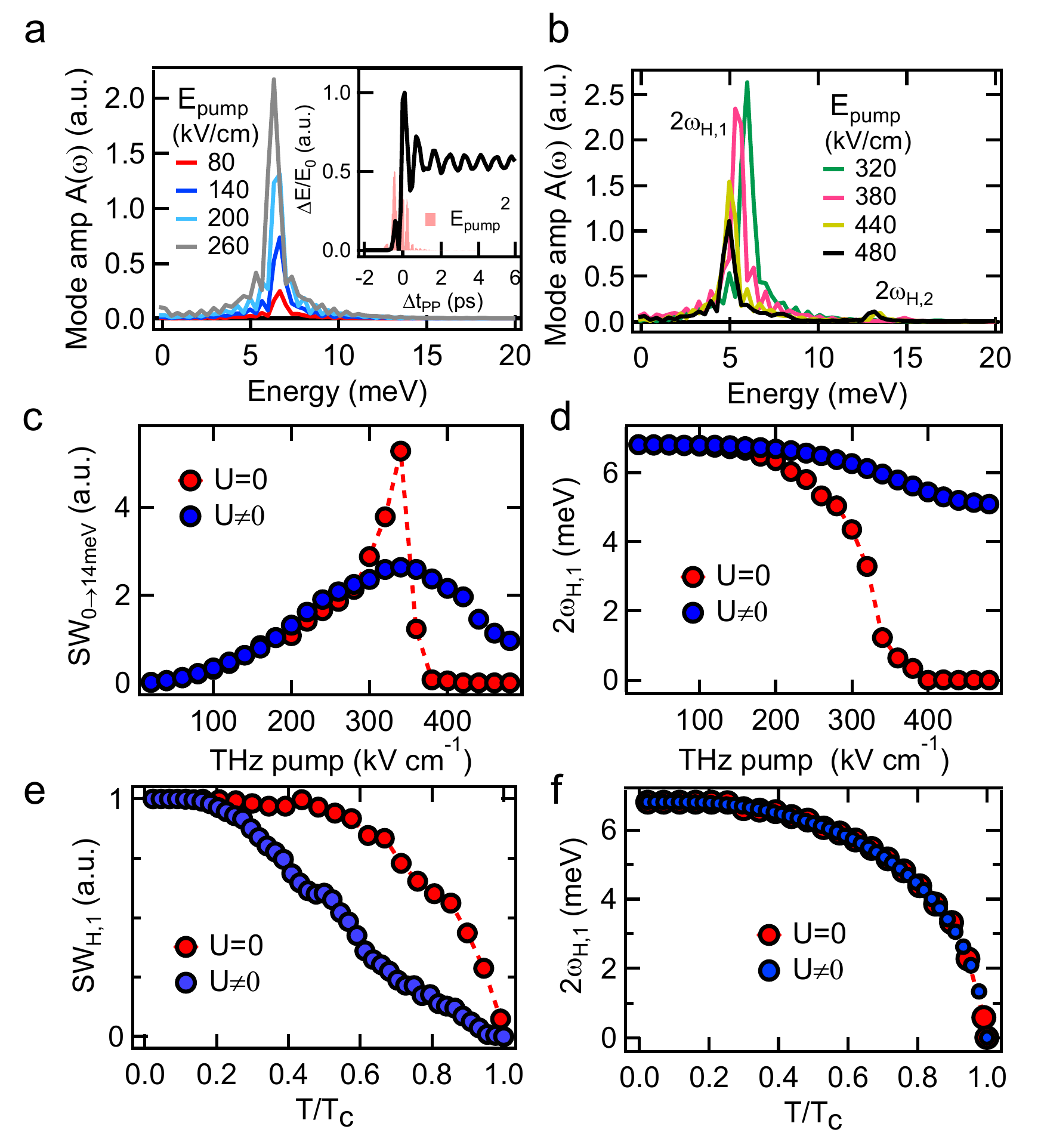}
	\caption{\textbf{Fig. 4. Gauge-invariant quantum kinetic calculation of the THz-driven hybrid Higgs dynamics.}  $\textbf {a},$ Calculated Higgs mode spectra for low pump $E$-field strengths. Inset: Calculated $\Delta E_{\mathrm{}}/E_{0}$ for peak E$_{\mathrm{pump}}$ = 380 kV/cm (black) and its comparison with the waveform E$_{\mathrm{pump}}^2$ (pink, shaded) of the applied THz pump pulse in the experiment. 
	$\textbf {b},$ Calculated Higgs mode spectra for higher field strengths show a decrease in amplitude and redshift of the Higgs mode $\omega_{H,1}$, consistent with experimentally measured coherent responses in Figs.~3c and 3d. Note that a second $\omega_{H,2}$ appears at higher $E_{\mathrm{pump}}$-field strengths and gets stronger at elevated $E_{\mathrm{pump}}$-field strengths. Although the $\omega_{H,2}$ mode is outside the experimental sampling width, it is revealed by the distinct nonlinear THz field dependence of spectral weight controlled by THz pump (Fig.~3d). 
	$\textbf {c},$ $E_{\mathrm{pump}}$-field dependence of the calculated Higgs mode $\omega_{H,1}$ spectral weight without inter-band coupling, i.e., $U=0$ (red circles) and with strong inter-band coupling $U\neq 0$ (blue circles). 
	$\textbf {d},$ Plot of the Higgs mode frequency $2\omega_{H,1}$ as a function of $E_{\mathrm{pump}}$-field strength without inter-band coupling (red circles) and for strong inter-band coupling (blue circles). 
	Our simulations for strong $U$ (blue circles) are in full agreement with the experimental results in Figs.~3c-3d and in sharp contrast with the one-band SC results obtained for $U=0$ (red circles). 
	$\textbf {e, f,}$ The calculated spectral weight SW$_{0\rightarrow14~\mathrm{meV}}$ \textbf{e} and resonance position \textbf{f} are plotted as a function of temperature for a fixed pump field of $56.0$~kV/cm  for $U = 0$ (red circles) and $U\neq 0$ (blue circles). With inter-band coupling, the SW is strongly suppressed, by about 60$\%$ up to a temperature of $0.6~T_C$, while at the same time the mode frequency is only slightly redshifted, by about 15$\%$, before a full spectral weight depletion is observable towards $T_C$. 
	These simulations are in agreement with the hybrid Higgs behavior in FeSCs and differ from one-band superconductors showing comparable change of SW and position of the Higgs mode with increasing temperature (red circles).}
	\label{Fig1}
\end{figure*}

To put the above 
Higgs mode findings on a rigorous footing, we calculate two--dimensional THz coherent nonlinear 
 spectra  by  extending the 
gauge-invariant density matrix equations of motion theory of Ref. \cite{mootz2020}, as outlined in the supplement (Section S6). 
Using the results of these calculations, 
we propose a physical mechanism  that explains the distinct differences of the Higgs mode resonance 
in the four--wave--mixing spectra between the strong and weak inter-band interaction limits. For this, we calculate the 
APS and quantum transport nonlinearities \cite{mootz2020} 
 driven by two intense phase--locked THz E-field pulses for an effective 3-pocket BCS model of FeSCs~\cite{fernandes:2016}. This model   includes both intraband and interband pairing interactions, as well as  asymmetry between electron and hole pockets. 
We thus calculate the nonlinear differential field transmission $\Delta E_{\mathrm{}}/E_{0}$ for  two phase--locked THz pulses,  which allows for a direct comparison of our theory with the experiment (supplementary). 

The inset of Fig.~4a presents the calculated $\Delta E/E_{0}$ (black line), 
shown together with E$_{\mathrm{pump}}^2(t)$ of the applied experimental pump pulse (pink shade).
The calculated Higgs mode spectra, Fig.~4a (regime I) and Fig.~4b (regime II), are dominated by a resonance close to 6.8~meV for low pump fluences, which corresponds to $\omega_{\mathrm{\mathrm{H},1}}$. This resonance grows up to pump fields $E_\mathrm{THz}\approx 320.0$~kV/cm for the parameters used here, with minimal redshift and without any significant SW at $\omega_{\mathrm{\mathrm{H},2}}$ (Fig.~4a). Interestingly, for higher fields (Fig.~4b), we obtain both a redshift and a decrease of the oscillation amplitude. In this regime II, SW emerges close to 15.0\,~meV, outside of our experimental bandwidth, in the  frequency regime of the  $\omega_{\mathrm{\mathrm{H},2}}$ Higgs mode. The latter mode is strongly suppressed due to damping induced via  electron-hole asymmetry  (supplementary). Specifically, the ellipticity of the e-pockets increases the DOS along the pump field direction and thus increases the damping of  mainly the $\omega_{\mathrm{\mathrm{H},2}}$ resonance, which leads to a transfer of oscillator strength to the continuum. This damping  has a much smaller influence on the $\omega_{\mathrm{\mathrm{H},1}}$ resonance  that arises largely from the hole pockets.
Most importantly, the strong  interband coupling expected in FeSCs leads to a decrease in the  $\omega_{\mathrm{\mathrm{H},1}}$ resonance amplitude,  with  SW reduction accompanied by a  persisting mode frequency
 This behavior of the  
multi--band model with strong $U$ is clearly seen in the raw experimental data.   
Note that, while in regime I  we observe an increase in the mode amplitude {\em without} any significant redshift, in  regime II, the decrease in $\omega_{\mathrm{\mathrm{H},1}}$ resonance is accompanied by a small redshift. This  behavior of the hybrid Higgs mode contradicts the one--band behavior, recovered by setting $U=0$, and is in excellent agreement with our experimental observations in the FeSC system (Figs. 3c-3e).

To scrutinize further the critical role of the  strong interband interaction $U$, we show the fluence dependence of the coherent Higgs SW close to $\omega_{\mathrm{\mathrm{H},1}}$
in Fig.~4c. SW$_{0\rightarrow14~\mathrm{meV}}$ differs markedly between the calculation with strong $U\neq 0$ (blue circles) and that without inter-pocket interaction $U=0$ (red circles), which 
resembles the one-band BCS quench results. Importantly, the Higgs mode SW$_{0\rightarrow14~\mathrm{meV}}$ for strong $U$ grows at low pump fluences (regime I), followed by a saturation and then decrease at elevated $E_\mathrm{pump}$ (regime II), consistent with the experiment. 
Meanwhile, Fig.~4d demonstrates that the resonance frequency remains constant in regime I, despite the strong increase of SW, and then redshifts in regime II, yet by much less than in the one--band system (compare $U\neq 0$ (blue circles) vs. $U=0$ (red circles)).  Without inter-pocket $U$ ($U=0$ , red circles in Figs.~4c-4d), the SW of Higgs mode $\omega_{\mathrm{\mathrm{H},1}}$ grows monotonically up to a  quench of roughly 90$\%$. A further increase of the pump field leads to a complete quench of the order parameter $\Delta_1$ and a decrease of the SW of Higgs mode $\omega_\mathrm{H,1}$ to zero, due to ransition from a damped oscillating Higgs phase to an exponential decay.  Based on the calculations in Figs. 4c-4d, the decrease of the spectral weight with interband coupling appears at a $\Delta_1$ quench close to 15$\%$, while without interband coupling the decrease of spectral weight is only observable close to the complete quench ($\sim$90$\%$) of the SC order parameter $\Delta_1$.  We conclude from this that  the spectral weight decrease at the lower Higgs resonance with low redshift is a direct consequence of the strong coupling between the electron and hole pockets
due to  large $U$. This $E_\mathrm{pump}$ dependence is the hallmark signature of the Higgs mode in FeSCs and is fully consistent 
with the Higgs mode behaviors observed experimentally. 

Finally, the temperature dependence of the hybrid Higgs mode predicted by our model is shown in Figs.~4e and 4f for $U = 0$ (red circles) and $U\neq 0$ (blue circles). With interband coupling, the SW is strongly suppressed, by about 60$\%$ up to a temperature of $0.6~T_C$, while at the same time the mode frequency is only slightly redshifted, by about 15$\%$, before a full spectral weight depletion is observable towards $T_C$. The strong suppression results from transfer of SW from mode $\omega_\mathrm{H,1}$ to the higher mode $\omega_\mathrm{H,2}$ with increasing temperature, since the higher SC gap $\Delta_2$ experiences stronger excitation by the applied pump $E^2$ with growing $T$. These simulations are in agreement with the hybrid Higgs behavior in Fig.~2 and differ from one-band superconductors showing comparable change of both SW and position of the Higgs mode with increasing temperature (red circles, Figs. 4e-4f). Moreover, our calculation without light-induced changes in the collective effects (only charge-density fluctuations) produces a significantly smaller $\Delta E_{\mathrm{}}/E_{0}$ signal in the non-perturbative excitation regime (Fig. S11, supplementary). Therefore, we conclude that the hybrid Higgs mode dominates over charge-density fluctuations in two-pulse coherent nonlinear signals in FeSCs, due to the different effects of the strong interband $U$ and multi-pocket bandstructure on QPs and on Higgs collective modes. 

In summary, we provide distinguishing features for Higgs modes and coherent excitations in FeSCs, which differ significantly from any previously observed collective mode in other superconducting materials: 2$\Delta_{\mathrm{SC}}$ amplitude oscillations displaying a robust mode resonance frequency position despite a large change of its spectral weight, more than 50 \%, on the THz electric field. This unusual nonlinear quantum behavior provides evidence for a compelling mechanism of hybrid Higgs from the two band entanglement in FeSCs. Our results also warrant further investigation of  Higgs collective modes  through broadband THz 2D spectroscopy in the coherent driven regime.

\bibliographystyle{Science}

\section*{Acknowledgments}

This work was supported by National Science Foundation 1905981 (THz spectroscopy and data analysis). 
The work at UW-Madison (synthesis and characterizations of epitaxial thin films) was supported by the US Department of Energy (DOE), Office of Science, Office of Basic Energy Sciences (BES), under award number DE-FG02-06ER46327.
Theory work at the University of Alabama, Birmingham was supported by the US Department of Energy under contract \# DE-SC0019137 (M.M and I.E.P) and was made possible in part by a grant for high performance computing resources and technical support from the Alabama Supercomputer Authority.

\section*{Author Contributions}
C.V., L.L. and X. Y. performed the THz spectroscopy measurements with J.W.’s supervision.
J.H.K, C.S. and C.B.E. grew the samples and performed crystalline quality and transport characterizations. M.M.
and I.E.P. developed the theory for the hybrid Higgs mode and performed calculations. Y. G. C and E. E.
H made Ba122 target for epitaxial thin films. J.W. and C.V. analyzed the THz data with the help of L.L., D.C., C.H., R.J.H.K. and Z.L. The paper is written by J.W., M.M., I.E.P. and C.V. with discussions from all authors. J.W. conceived and supervised the project.

\section*{Correspondence} Correspondence and requests for materials should be addressed to J.W. (jwang@ameslab.gov;
jgwang@iastate.edu).


\clearpage


\end{document}